\documentstyle[11pt,twoside,colloqOHP,epsfig]{article}
\begin{document}
\title{Hot Jupiters:  Lands of Plenty}
\author{David Charbonneau}
\affil{Harvard-Smithsonian Center for Astrophysics, 60 Garden Street, Cambridge, MA 02138 USA [dcharbonneau@cfa.harvard.edu]} 
\begin{abstract}
In late August 2005, 80 researchers from more than 15 countries convened for a 4-day conference
entitled ``The Tenth Anniversary of 51~Peg~b:  Status and Prospects for Hot Jupiter Studies''.
The meeting was held at l'Observatoire de Haute-Provence, the location of the 1.93-m
telescope and ELODIE spectrograph used to discover the planetary companion to 51~Peg
roughly 10 years ago.  I summarize several dominant themes that emerged from the
meeting, including (i) recent improvements in the precision of radial velocity measurements
of nearby, Sun-like stars, (ii) the continued value of individual, newly-discovered
planets of novel character to expand the parameter space with which the theory must contend, and
(iii) the crucial role of space-based observatories in efforts to characterize hot Jupiter planets.
I also present the returns of an informal poll of the conference attendees
conducted on the last day of the meeting, which may be amusing to revisit a decade hence.  
\end{abstract}
\section{Dominant Themes from the Meeting}
Prior to the October 1995 announcement of the discovery of the planetary companion to
51~Peg (Mayor \& Queloz 1995), few astronomers foresaw the existence of hot Jupiter
exoplanets (for a notable exception, see Struve 1952).  In the past decade, this subclass
of exoplanets has been by far the best observationally studied, due to the rich set of follow-up
techniques enabled by the proximity of the planet to the host star.  The entire volume 
of this conference proceedings is replete with descriptions of dozens of exciting studies,
some in progress, others planned for the near-future, detailing a wide diversity
of observational efforts too numerous to list here.  Radial velocity observations
remain the dominant tool for the detection of extrasolar planets, and provide precise
estimates of the orbital parameters of these objects.  However, these measurements
alone yield little information about the planetary body directly (other than
the value of the minimum mass).  Once combined with complementary 
information, a rich, observationally-constrained picture of these objects emerges.  
An  abbreviated list of such measurements include 
observations of the planetary photometric transits, 
reflected starlight, thermal emission, atmospheres and exospheres via
transmission spectroscopy, as well as observations of the central star including its
metallicity, magnetic activity, and the degree of alignment between the orbital and stellar
rotational axes.  While the breadth of observational and theoretical work 
presented at the meeting is far too great to be justly summarized here,
I describe below several important themes that emerged over the course
of the conference.

\subsection{Impressive Leaps in Radial Velocity Precision}
A significant development in the past couple years is that a radial
velocity precision of $3-10 \ {\rm m \,s^{-1}}$, previously
achieved by only a few teams, has now been put to wide-spread use
by numerous groups around the globe.  The benefit is that the additional
researchers and telescopes now monitor a much more diverse set of primary stars than
the F, G, K, and early-M dwarfs that have been the 
principal targets over the past decade (e.g. Marcy et al.\ 2005; Udry et al.\ 2006).
Several examples of the parameter spaces under 
investigation are (1) targeted searches for hot Jupiters (Fischer et al.\ 2005; Bouchy et al.\ 2005), 
(2) surveys targeting only low-mass stars (Bonfils et al.\ 2005; Endl et al.\ 2003),
(3) surveys dedicated to monitoring binary stars (Mart{\'{\i}}nez Fiorenzano et al.\ 2005; Konacki 2005),
(4) searches for planets orbiting young stars (Esposito et al.\ 2005), and,
(5) searches for planets orbiting evolved stars (Hatzes et al.\ 2005; Sato et al.\ 2005a).
These numerous efforts promise a rich catch of planetary systems that will
flush out the full parameter space of planet formation.  
The forefront of Doppler precision is now well below the level of $3 \ {\rm m \, s^{-1}}$:
The HARPS instrument has yielded a precision of $1 \ {\rm m \, s^{-1}}$ (Santos et al.\ 2004),
and data presented at the conference hinted that further improvements
are close at hand (e.g. Mayor et al.\ 2005).  A precision of tens of ${\rm cm \, s^{-1}}$
would enable the mass determination of terrestrial planets orbiting within
the habitable zone of low-mass stars, a very exciting prospect indeed.

\subsection{The Ongoing Value of Individual Objects}
Despite the benefits of a statistical analysis of the hot Jupiter population
as a whole, the detection of an individual planet of novel character 
can still significantly impact the field.  Consider the
following recent examples:  (1) The planet orbiting the star HD~149026 (Sato et al.\ 2005b) 
has a radius significantly less than Jupiter ($0.726 \pm 0.064 \ R_{\rm Jup}$; Charbonneau et al.\ 2006),
indicating the presence of a large core of solid material.  This would seemingly
prove that this planet formed through core accretion, as opposed to gravitational
instability.  (2) The planet orbiting one member of the stellar triple HD~188753
presents, in turn, a challenge to the core-accretion model, 
since the binary companion would have likely truncated
the protoplanetary disk to a radius less than 1.3~AU, which lies interior to the snow line.  Recent work 
(Pfahl 2005) suggests that this system may have formed in the dense environment of a
stellar cluster.  (3) Two new hot Neptunes (HD~4308b; Udry et al.\ 2005, and GJ 581b; Bonfils et al.\ 2005)
were the welcome news with which the conference began.  The presence of 
such objects and a handful of others (Butler et al.\ 2004; McArthur et al. 2004;
Santos et al. 2004; Rivera et al.\ 2005) 
may hint at a large population of low-mass planetary companions in short-period orbits.
The detection of the first transiting Neptune-mass planet is eagerly awaited.
(4) Soon after the conference ended, the announcement
of the discovery of the transiting planet HD~189733~b (Bouchy et al.\ 2005) left many observers
scrambling to gather data both from the ground and space.  This enthusiasm
was motivated by both the proximity of the system ($d=19$\, pc, making it the closest
known transiting exoplanet), and the 
favorable ratio of the planet's area to that of the star, both of which facilitate
observations geared to detect the planet directly in either emitted or 
reflected light.

\subsection{The Need for Dynamical Mass Estimates}
The controversy over the mass determinations (and hence planetary status)
for the recently imaged companions to 2MASS1207 (Chauvin et al.\ 2005) 
and GQ~Lup (Neuh{\"a}user et al.\ 2005) arises primarily
from the lack of direct constraints on the theoretical emission
models, upon which the mass estimates are based.  The most
reliable means to resolve this issue would be to locate similar
objects in systems for which the masses may be determined dynamically,
and thus provide the strict (and unforgiving) constraints on
such models, as is currently ongoing for M-dwarfs (e.g. Ribas 2005).
Indeed the greatest asset of transiting exoplanets is that the
masses and radii may be determined robustly.

\subsection{The Crucial Role of Space-Based Observatories}
One of the most interesting aspects of hot Jupiter exoplanets
is the set of opportunities that these objects afford for direct study.
In that regard, it is very important to note the pivotal contributions
from several space-based observatories, some of which where 
designed before such planets (and hence observations
of these objects) could have been foreseen.  Consider the numerous attempts
to study the atmospheres and exospheres of transiting hot Jupiter planets:
A host of ground-based efforts (e.g. Brown et al.\ 2002; Bundy \& Marcy 2000; Deming et al.\ 2005a;
Moutou et al.\ 2001, 2003; Narita et al.\ 2005; Winn et al.\ 2004) have yielded only upper limits (albeit
very useful ones), whereas the detections have all come from the Hubble
Space Telescope (Charbonneau et al.\ 2002; Vidal-Madjar et al.\ 2003).
Similarly, upper limits from ground-based attempts to measure the thermal
emission from such planets (e.g. Richardson et al.\ 2003a, 2003b) have 
recently been met by robust detections with the Spitzer Space Telescope
(Charbonneau et al.\ 2005; Deming et al 2005b; for a comparison to theoretical
models, see Burrows et al.\ 2005; Fortney et al.\ 2005; Seager et al.\ 2005).  And, despite remarkable
ground-based photometry of transit curves (e.g. Moutou et al.\ 2004;
Charbonneau et al.\ 2006; Holman et al.\ 2006), no such efforts have approached
the exquisite results from both the STIS and ACS instruments aboard the
Hubble Space Telescope (e.g. Brown et al.\ 2001, 2005).  The MOST
satellite (Walker et al.\ 2003) will either detect reflected light
from one or more hot Jupiters, or place very stringent upper limits
on the albedos (Walker et al.\ 2005), perhaps finally bringing respite
to frustrated ground-based searches for this signal (Leigh et al.\ 2003a, 2003b; Collier-Cameron et al.\ 2002;
Charbonneau et al.\ 1999).

\section{A Poll of Conference Attendees}
The rapid pace of successes over the past decade have inspired many
fond remembrances of the prevailing wisdom prior to October 1995, some of which
may be more accurate than others.  In advance preparation for the conference
celebrating the twentieth anniversary of the discovery of 51 Peg b (to be held
in 2015), I put four questions to the attendees in the final
session of the meeting, on the afternoon of August 25th, 2005.  In these
questions, the participants were asked to speculate as to the time
scale and technical methods of significant future advances in the field
of exoplanet science.  The purpose of this survey was to record informally
the opinions of the conference attendees as to these important questions,
so that the replies could be revisited for both interest and enjoyment
a decade hence.  In order to avoid burdening the questions with unduly
technical and lengthy definitions, the precise meaning of certain
phrases (such as ``habitable zone'' and ``extraterrestrial life'') was left somewhat
vague:  I note that this ambiguity pervades the more rigorous
(and refereed) literature as well.   The questions, and the tabulated responses, are
presented in Figures 1 \& 2.

\begin{figure}
\plotone{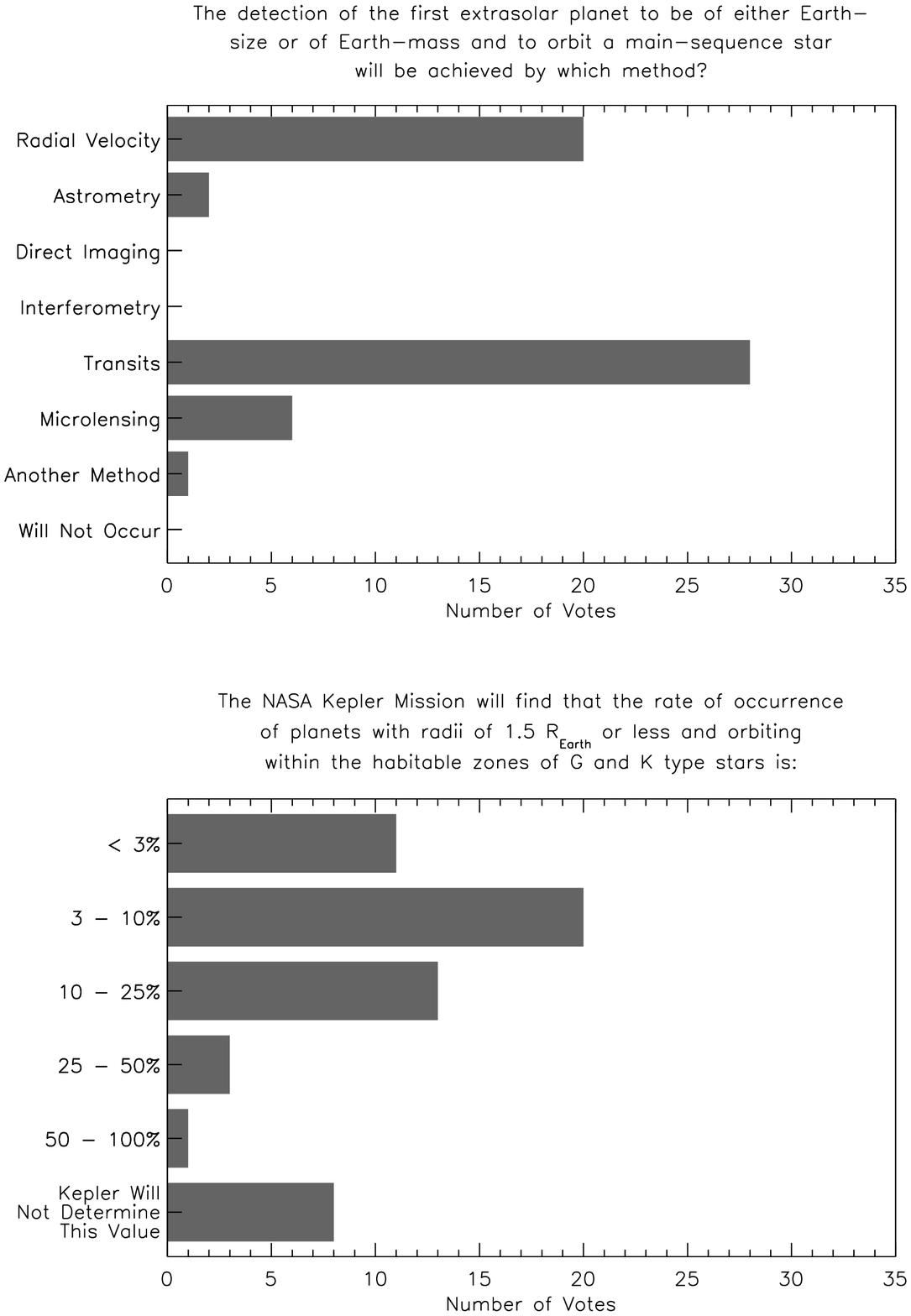}
\caption{Conference participants who attended the final
session were asked to vote on four speculative questions (see Fig.\ 2).
\emph{Upper panel:} Attendees favored either the transit method or
the Doppler technique to detect the first Earth-like planet.
\emph{Lower panel:} Attendees were divided as to the rate of
occurrence of Earth-like planets as measured by the Kepler Mission, but few thought it would
be greater than 25\%.}
\end{figure}

\begin{figure}
\plotone{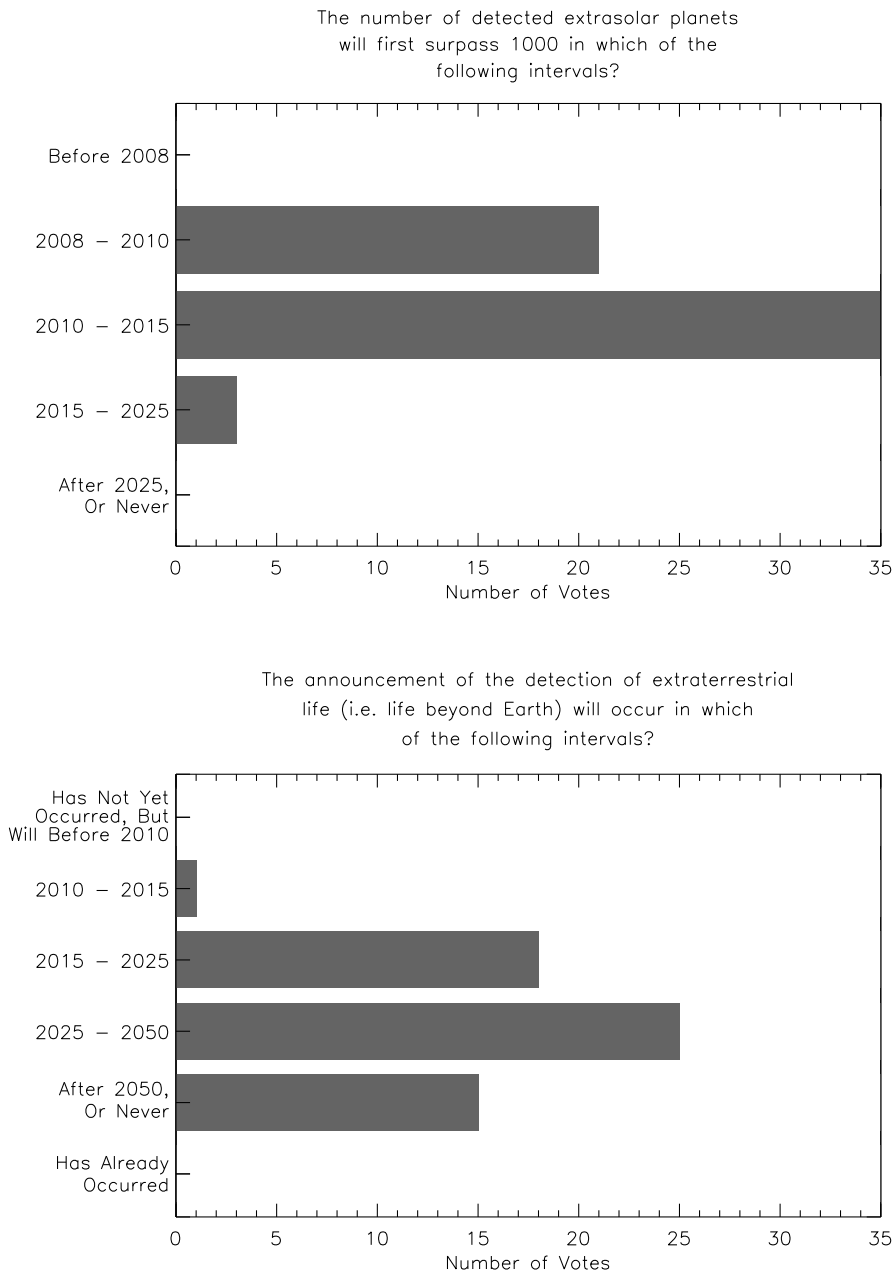}
\caption{\emph{Upper panel:} Roughly 160 extrasolar planets have been detected in the past decade.
Most participants felt that rate of detection would be greater in the second
decade, with 1000 such planets detected by 2015, 10 years hence.  \emph{Lower panel:} All but 
one of the attendees thought that extraterrestrial life would not be detected in the next decade, but a 
majority felt it would occur prior to 2050.}
\end{figure}

A quick summary of the responses is as follows.  Most participants felt that 
the transit method and/or the radial velocity method would yield the first
detection of an Earth-like planet.  Notably, no respondents felt that
this would first be accomplished by interferometry or direct
imaging (such as coronagraphy), efforts which currently receive a great deal of 
support.  (It must be noted, of course, that missions such as TPF and
Darwin are geared toward the spectral characterization of such
planets, not simply their discovery.)  Attendees were divided as to 
the value that the NASA Kepler Mission will determine for the
rate of occurrence of Earth-like planets in the habitable zones of
Sun-like stars.  Few thought it would
be greater than 25\%, and a notable fraction questioned the mission's
ability to determine this value at all.  Whatever the method of discovery,
nearly all participants felt that the number of detected exoplanets
would exceed 1000 by the year 2015.  As for the big question of
extraterrestrial life, all but one of the attendees
felt that its discovery would not occur by that date.
A majority voted that its detection would be achieved prior to 2050, but
a significant number felt it would occur after that year, or not at all.

With the exciting discoveries of the previous decade as our guide, we
can only assume that the prevailing wisdom will, once again, be 
proven wholly unjustified.

\end{document}